\documentclass[3p,times]{elsarticle}
\usepackage{ecrc}
\usepackage{url} 
\runauth{}
\volume{00}
\firstpage{1}
\journalname{AI Open}

\runauth{}

\jid{procs}

\jnltitlelogo{AI Open}
\usepackage{amssymb}
\usepackage[figuresright]{rotating}
\usepackage[misc,geometry]{ifsym}
\usepackage{footmisc}
\CopyrightLine{2011}{Published by Elsevier Ltd.}

\begin{document}
\begin{frontmatter}
\dochead{}

\title{Information Retrieval Meets Large Language Models: A Strategic Report from Chinese IR Community}


\author{{{Qingyao AI}$^a$}{,}\hspace{0.25em}
{{Ting BAI}$^b$}{,}\hspace{0.25em}
{{Zhao CAO}$^c$}{,}\hspace{0.25em}
{{Yi CHANG}$^d$}{,}\hspace{0.25em}
{{Jiawei CHEN}(\Letter)$^e$\footnotetext[1]{Email: sleepyhunt@zju.edu.cn; xiangnanhe@gmail.com; pzhang@tju.edu.cn}}{,}\hspace{0.25em}
{{Zhumin CHEN}$^f$}{,}\hspace{0.25em}
{{Zhiyong CHENG}$^g$}{,}\hspace{0.25em}
{{Shoubin DONG}$^h$}{,}\hspace{0.25em}
{{Zhicheng DOU}$^i$}{,}\hspace{0.25em}
{{Fuli FENG}$^j$}{,}\hspace{0.25em}
{{Shen GAO}$^f$}{,}\hspace{0.25em}
{{Jiafeng GUO}$^k$}{,}\hspace{0.25em}
{{Xiangnan HE}(\Letter)$^j$}{,}\hspace{0.25em}
{{Yanyan LAN}$^a$}{,}\hspace{0.25em}
{{Chenliang LI}$^l$}{,}\hspace{0.25em}
{{Yiqun LIU}$^a$}{,}\hspace{0.25em}
{{Ziyu LYU}$^m$}{,}\hspace{0.25em}
{{Weizhi MA}$^a$}{,}\hspace{0.25em}
{{Jun MA}$^f$}{,}\hspace{0.25em}
{{Zhaochun REN}$^f$}{,}\hspace{0.25em}
{{Pengjie REN}$^f$}{,}\hspace{0.25em}
{{Zhiqiang WANG}$^n$}{,}\hspace{0.25em}
{{Mingwen WANG}$^o$}{,}\hspace{0.25em}
{{Ji-Rong WEN}$^i$}{,}\hspace{0.25em}
{{Le WU}$^p$}{,}\hspace{0.25em}
{{Xin XIN}$^f$}{,}\hspace{0.25em}
{{Jun XU}$^i$}{,}\hspace{0.25em}
{{Dawei YIN}$^q$}{,}\hspace{0.25em}
{{Peng ZHANG}(\Letter)$^r$}{,}\hspace{0.25em}
{{Fan ZHANG}$^l$}{,}\hspace{0.25em}
{{Weinan ZHANG}$^s$}{,}\hspace{0.25em}
{{Min ZHANG}$^a$}{,}\hspace{0.25em}
{{Xiaofei ZHU}$^t$}\hspace{0.25em}
\\
\hspace*{\fill} 
\\
  \textsuperscript{a}\textit{Tsinghua University}{,}\hspace{0.25em}
  \textsuperscript{b}\textit{Beijing University of Posts and Telecommunications}{,}\hspace{0.25em}
  \textsuperscript{c}\textit{Huawei Technologies Ltd. Co}{,}\hspace{0.25em} \textsuperscript{d}\textit{Jilin University}{,}\hspace{0.25em}\textsuperscript{e}\textit{Zhejiang University}{,}\hspace{0.25em} \textsuperscript{f}\textit{Shandong University}{,}\hspace{0.25em} \textsuperscript{g}\textit{Shandong Artificial Intelligence Institute}{,}\hspace{0.25em} \textsuperscript{h}\textit{South China University of Technology}{,}\hspace{0.25em} \textsuperscript{i}\textit{Renmin University of China}{,}\hspace{0.25em} \textsuperscript{j}\textit{University of Science and Technology of China}{,}\hspace{0.25em} \textsuperscript{k}\textit{Institute of Computing Technology, Chinese Academy of Sciences}{,}\hspace{0.25em} \textsuperscript{l}\textit{Wuhan University}{,}\hspace{0.25em} \textsuperscript{m}\textit{Shenzhen Institute of Advanced Technology, Chinese Academy of Sciences}{,}\hspace{0.25em} \textsuperscript{n}\textit{Shanxi University}{,}\hspace{0.25em} \textsuperscript{o}\textit{Jiangxi Normal University}{,}\hspace{0.25em} \textsuperscript{p}\textit{Hefei University of Technology}{,}\hspace{0.25em} \textsuperscript{q}\textit{Baidu Inc.}{,}\hspace{0.25em} \textsuperscript{r}\textit{Tianjin University}{,}\hspace{0.25em} \textsuperscript{s}\textit{Shanghai Jiao Tong University}{,}\hspace{0.25em} \textsuperscript{t}\textit{Chongqing University of Technology}\hspace{0.25em}
}

\begin{abstract}
The research field of Information Retrieval (IR) has evolved significantly, expanding beyond traditional search to meet diverse user information needs. Recently, Large Language Models (LLMs) have demonstrated exceptional capabilities in text understanding, generation, and knowledge inference, opening up exciting avenues for IR research. LLMs not only facilitate generative retrieval but also offer improved solutions for user understanding, model evaluation, and user-system interactions. 
More importantly, the synergistic relationship among IR models, LLMs, and humans forms a new technical paradigm that is more powerful for information seeking. IR models provide real-time and relevant information, LLMs contribute internal knowledge, and humans play a central role of demanders and evaluators to the reliability of information services. Nevertheless, significant challenges exist, including computational costs, credibility concerns, domain-specific limitations, and ethical considerations. To thoroughly discuss the transformative impact of LLMs on IR research, the Chinese IR community conducted a strategic workshop in April 2023, yielding valuable insights. This paper provides a summary of the workshop's outcomes, including the rethinking of IR's core values, the mutual enhancement of LLMs and IR, the proposal of a novel IR technical paradigm, and open challenges.

\end{abstract}


\end{frontmatter}


\section{Introduction}

In the past few decades, Information Retrieval (IR) has experienced significant growth and development in both industry and academia. In early stage, IR research mainly focused on search, which aimed to assist users in finding relevant information~\cite{kobayashi2000information}. In recent years, the scope of IR research has expanded to encompass a wide range of online applications and scenarios~\cite{manoj2008information}. This diversification is evident in the rise of recommendation systems as a prominent research area, as observed in flagship conferences ACM SIGIR from 2018 to 2023. Additionally, exciting research directions such as conversational systems, user modeling, and knowledge extraction, have also emerged~\cite{faggioli2021hierarchical, li2022user}. These advancements reflect an evolution in the core value of IR, moving beyond merely retrieving relevant documents to meeting the information needs of users~\cite{Chen2020InformationRA}.


Large Language Models (LLMs) have demonstrated remarkable capabilities in various aspects of text understanding, generation, knowledge inference, and compositional generalization~\cite{bubeck2023sparks,liu2023summary}. As a result, LLMs have the potential to open up new research directions in the field of IR. These include, but are not limited to: 
\begin{itemize}
    \item Enabling IR systems to generate content directly that satisfies user information needs, known as generative retrieval~\cite{lee2022generative} and generative recommendation~\cite{wang2023generative}. An example of this is the New Bing\footnote{\url{https://www.bing.com/new}}, which allows for content generation to meet user queries.
    \item Enhancing the understanding of user intents and behaviors by incorporating rich contextual information into IR system.
    \item Creating opportunities for the development of superior indexing systems that can handle dynamic, semantically-aware, and multi-modal data.
    \item Providing better methods for evaluating model accuracy and interpretability in IR tasks.
    \item  Facilitating enhanced interactive experiences between users and IR systems. These advancements highlight the potential of LLMs to revolutionize various aspects of IR research and practice.
\end{itemize}

\begin{figure*}
    \centering
    \includegraphics[width=0.7\linewidth]{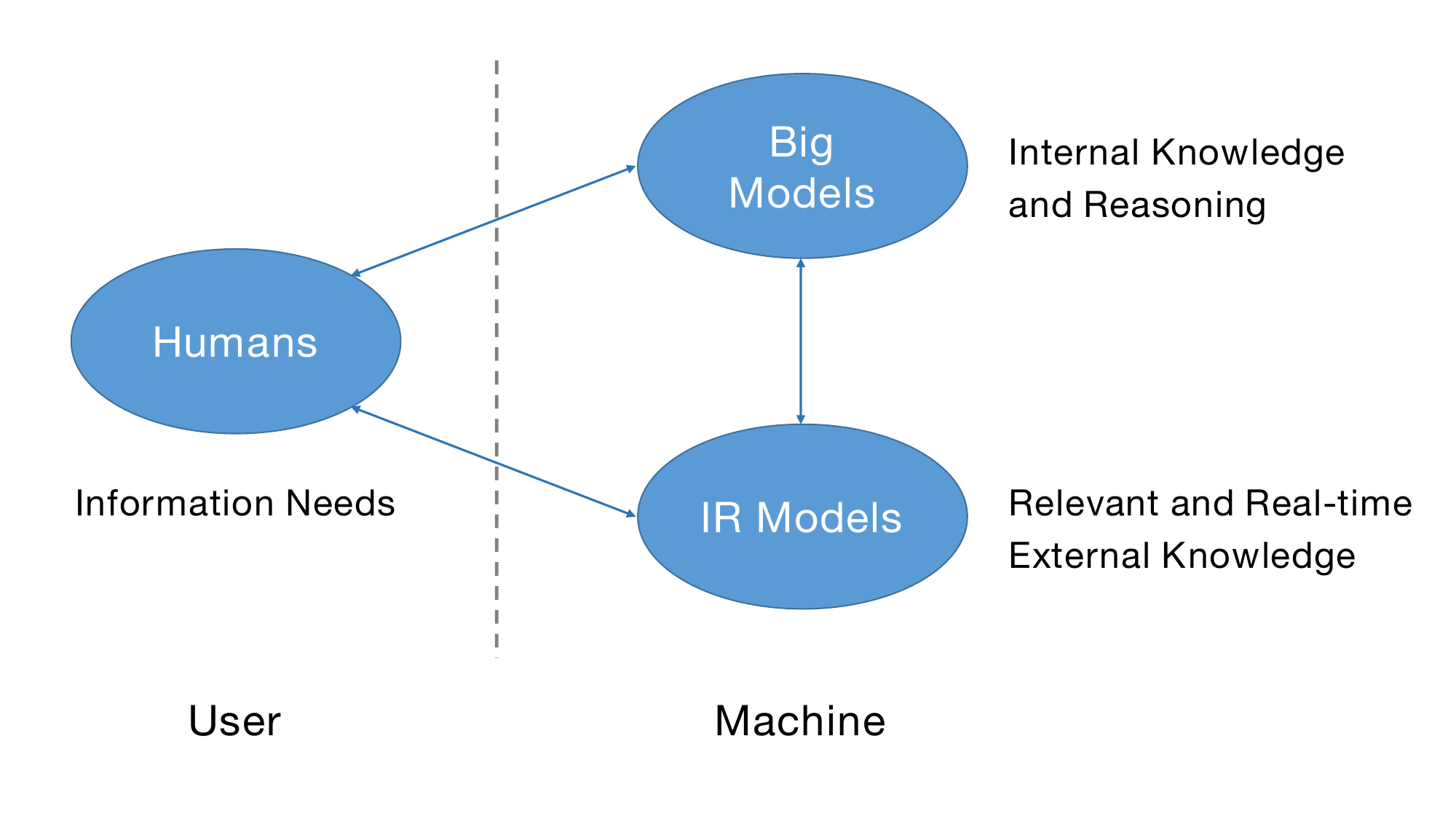}
    \caption{Our proposed new IR technical paradigm introduces three key elements: Humans, IR models, and LLMs. The synergistic relationship among them not only facilitates mutual enhancement but also enables the fulfillment of new tasks as an organic whole. }
    \label{fig:1}
\end{figure*}

Reciprocally, IR technology plays a crucial role in supporting the development of LLMs~\cite{zheng2023bimgpt, jeronymo2023inparsv2}. 
The interaction among humans, IR models, and LLMs forms a new triangular IR paradigm, as depicted in Figure~\ref{fig:1}. In this paradigm, the IR model acts as the means to acquire external knowledge, offering real-time, up-to-date, and relevant information to both LLMs and humans. 
LLMs contribute precise internal knowledge and information, leveraging their powerful reasoning abilities to provide high-quality responses. 
Humans, in their role as demanders and evaluators, play a central role in the retrieval process.
It is important to emphasize that LLMs, without the inclusion of IR models, possess limitations in terms of short-term memory and reasoning. 
These limitations could result in the lack of effective integration of new and existing knowledge, hindering LLMs' ability to provide the information that humans need. 
The IR model, functioning as a ``knowledge base'' equipped with matching and retrieval capabilities, addresses these limitations by compensating for the lack of factual consistency and long-term memory in LLMs. This combination of ``knowledge+reasoning'' establishes a dual-wheel drive, enabling a more intelligent and reliable information service system for humans~\cite{kim2022ask, zheng2023bimgpt}.


While LLMs for IR hold promise, they also face significant challenges that cannot be ignored:
\begin{itemize}
\item High computational costs: LLMs for IR require substantial computational resources, which can limit their use and research by small and medium-sized companies and institutions. Furthermore, deploying models on terminal devices and ensuring real-time performance can be challenging~\cite{edalati2021kronecker}.
\item Credibility concerns: The credibility of LLMs for IR is relatively low, as they have been found to provide misleading explanations, incorrect answers, and unreliable information sources. This can undermine people's trust in LLMs~\cite{liu2023evaluating}.
\item Performance limitations in specific fields: LLMs may struggle to perform well in specific domains due to the lack of high-quality datasets, differences in data representation, and other related issues. The limited domain-specific knowledge hampers the rapid implementation and application of LLMs~\cite{nori2023capabilities,liu2023deidgpt}.
\item Ethical and moral considerations: The emergence of generative IR systems raises the requirements for ethical and moral evaluation standards. Ensuring fairness, impartiality, and ethics in the generated content is a significant challenge. It is inherently difficult to guarantee that LLMs meet these regulatory and ethical requirements~\cite{zhang2023chatgpt,blairstanek2023gpt3}.
\end{itemize}


Given the flourishing research in the IR field and the promising opportunities brought about by LLMs, it is a favorable moment to envision the future advancements in IR. To this end, the Chinese IR community organized a strategic workshop on April 14th and 15th, 2023, aiming to explore future opportunities and challenges. This paper presents the key findings and conclusions drawn from the workshop, including the rethinking of the core values of IR in Section~\ref{Sec:rethink_IR}, a discussion of how LLMs and IR can mutually enhance each other in Section~\ref{Sec:big_model_IR} and Section~\ref{Sec:IR_for_big_model}, the new IR paradigm of fusing humans, LLMs, and IR in Section~\ref{Sec:big_model+IR}, and a discussion of open challenges and future directions in Section ~\ref{Sec:challenges_discussions}.

\section{Rethinking Core Values of IR}
\label{Sec:rethink_IR}
Before delving into the effects of LLMs on IR research, it is crucial to rethink the IR discipline itself. This contemplation leads us to ponder the following questions: Firstly, what is the fundamental contribution of IR, as a scientific discipline, to other domains? Secondly, considering its extensive development over the past decades, what are the boundaries and extensions of IR? These questions are intended to promote comprehension of the fundamental principles of IR, and foster the integration of LLMs into IR while preserving its essence.

\subsection{Scientific Connotation of IR}
Classical IR concerns the retrieval of the information that satisfies a user's information need from the target corpus. Therefore, user, ranking, and corpus are the fundamental concepts in the IR discipline. In the light of this, it is important to re-examine the scientific connotation of IR from the three aspects.

\paragraph{User Perspective} Users are the core part of IR~\cite{manning:2008:introduction,Mei:2000:IRWeb}, since an IR system is supposed to serve users by satisfying their information needs.  
Therefore, user understanding is a fundamental problem in IR, and the community has made decades-long progress in user understanding. The studies of user understanding in IR mainly concentrate on two sides, respectively user intent understanding and user behavior modeling. 
\begin{itemize}
    \item User intent understanding plays a central role in IR. Commercial search engines like Google \footnote{\url{https://www.google.com}} and Bing \footnote{\url{https://www.bing.com}}, have evolved from keyword semantic matching to user intent optimization in recent decades. Understanding user intent is the first step for better user experiences, which has been broadly studied in various IR applications, e.g., web search \cite{Jansen:2007:UserIntentWeb,Teevan:2008:UserIntentWeb}, product search in e-commerce \cite{Su:2018:Product}, community question answering \cite{Chen:2012:CQA}, etc. 
    \item User behavior modeling is to comprehensively understand and model user behaviors when interacting with IR systems. IR community has studied diverse and multi-dimensional user behaviors including user interaction behaviors \cite{Agichtein:2006:userBehavior,Manavoglu:2003:userBehavior}, sequential behaviors \cite{Qi:2019:Sequential,Yuan:2020:Sequential}, mobility behaviors \cite{Yuan:2012:Mobility,zheng:2010:geolife} and social behaviors \cite{Long:2013:Social}. User behavior modeling helps to inject personalization into IR systems.
\end{itemize}

The techniques of user understanding in IR have provided valuable insights for other research domains. In natural language processing (NLP), 
 user intent understanding is the preliminary step to make further actions in task-oriented dialogue systems \cite{Alfieri:2022:Task}. User profiling and personalization are also utilized to develop persona-based dialogue systems \cite{Li:2016:Persona,Zhong:2020:Persona}. In social computing, user modeling is an important area in the design of personalized incentive mechanisms for encouraging participation \cite{Vassileva:2012:social}. 

\paragraph{Ranking Perspective}
Ranking lies at the core part of IR research by providing  an ordering of items towards user's information need \cite{Guo:2020:Ranking,Trabelsi:2021:Ranking}.  The formats of items are various, including but not limited to textual documents, images, and videos. 
The problem is usually formulated as ranking a collection of items that match the given query or context, according to some criterion like relevance and recency. Most IR systems are supported by a ranking module.

Different techniques have been used for constructing ranking models, from traditional ranking models to machine learning methods. The traditional ranking models include vector space models and probabilistic models (e.g. BM25 \cite{Robertson:2009:BM25}). The machine learning methods include the earliest learning to rank models (e.g. Rankboost \cite{Freund:2003:Rankboost} and LambdaMart \cite{Burges:2010:LambdaMART}) and the recent neural ranking models such as Deep Structured Semantic Models (DSSM) \cite{Huang:2013:DSSM} and DeepRank \cite{Liang:2017:DeepRank}. Ranking models have been extensively exploited in many IR applications such as ad-hoc retrieval \cite{Guo:2016:Ad-hoc,Jiang:2020:Ad-hoc,Fujiwara:2013:Ad-hoc}, recommender systems \cite{Karatzoglou:2013:RecSys,Vargas:2011:RecSys}, community question answering \cite{Kai:Zhang:CQA,Liu:2013:CQArank}, etc. 

It's worth noting that ranking is not restricted to IR alone, as it also has applications in a diverse range of other research fields.
For example, ChatGPT \cite{Long:2022:ChatGPT} collects comparison samples and utilizes a ranking labeler to train the reward model when aligning with user intents. In computational biology, learning-to-rank algorithms have been exploited to rank the candidate 3-D structures in protein structure prediction \cite{Duh:2008:Bio}.

\paragraph{Corpus Perspective}
For the corpus perspective, the connotation of IR lies in conducting effective search efficiently from a large volume of multi-modal information corpus \cite{kobayashi2000information}. Such connotation mainly lies in the following two folds:
\begin{itemize}
    \item Effective corpus representation. Conventional IR systems represent the documents with bag-of-words or TF-IDF methods \cite{kobayashi2000information}. Deep learning methods (e.g., dense retrieval) empowers the IR system to represent multi-modal corpus such as images and videos with dense latent vectors, which has largely improved the retrieval results and broadened the application scope of IR \cite{karpukhin2020dense,zhen2019deep}. Recently, contrastive learning  and self-supervised learning further improve the learning of corpus representation \cite{izacard2021towards}. 
    \item Efficient retrieval infrastructure. The infrastructure of IR involves fundamental facilities to construct an IR system, including but not limited to building dictionaries, indexing, scoring, distributed search, and caching \cite{manning:2008:introduction}. Such infrastructure determines how to organize the information source and thus largely affect the efficiency of IR \cite{manning:2008:introduction}. LLMs and generative IR \cite{tay2022transformer} shed new lights to reform the retrieval infrastructure.   
\end{itemize}

\subsection{Boundaries and Extensions of IR}
Although the growth of Web and mobile technology has greatly advanced the development of IR, it also results in some limitations. 
A common misconception is that IR simply involves ranking a set of objects in cyberspace, which is an inaccurate and narrow view.
As such, it is crucial to reconsider the boundaries and extensions of IR from various perspectives, including the input, process, output, and environment.

\paragraph{Ranking or Generation?}
The primary focus of classical IR pertains to the ordering of items based on the information need, such as the query in search engines or the context of recent interactions in recommender systems.
The advancement of large-scale generative models such as GPT~\cite{brown2020language} and Diffusion Models~\cite{rombach2022high} has expanded the content space, allowing exploration beyond existing sets of content through generation. Integrating the generation channel for information seeking opens up new paradigms in IR systems, such as generative search and recommendation~\cite{wang2023generative}. However, this extension presents challenges to building IR systems in infrastructure, algorithm, evaluation, \textit{etc}. For instance, it calls for the infrastructure to accommodate generative models at a large scale and distribute the content generated on the fly in an efficient manner, especially for multi-media content such as micro-videos. Additionally, it calls for algorithm innovation on the alignment of generative models to IR tasks, the aggregation of ranking and generation results, and the joint optimization of ranking and generation models. Furthermore, evaluation techniques must also be extended to accommodate unlimited content and emphasize new perspectives such as fidelity checking.

\paragraph{For Human or For Machine?} 
IR systems have been designed to assist \textit{human} users to access information resources. However, with the increasing prevalence of AI models, particularly LLMs, the user base of IR systems could potentially be expanded to include intelligent \textit{machines}. 
This has led to the emergence of a growing body of research referred to as IR-augmented methods or retrieval-enhanced machine learning (REML) \cite{Zamani2022REML} in various communities such as computer vision\cite{gur-etal-2021-cross-modal}, natural language processing \cite{DBLP:journals/corr/abs-2112-04426}, and machine learning\cite{10.5555/3045390.3045585}. 
From a conceptual perspective, IR systems can be regarded as a memory and access system for external data, information, or knowledge, capable of serving as a generic supporting technique for either humans or machines. Nevertheless, the shift from human to machine poses a multitude of research challenges throughout the entire process of the IR system, including indexing, representation, retrieval, ranking, and feedback. Additionally, there are many research issues to be addressed regarding the learning, evaluation, and deployment of IR technology in conjunction with AI models. Despite these challenges, it is encouraging to anticipate that IR could become a ubiquitous component of the AI paradigm in the future, providing valuable support to society at large.

\paragraph{For Finding or For Decision?} IR systems have long benefited people in finding desired information. For the extension of IR, the system should also support complicated, explainable, and long-term information seeking to help users with reasonable decision-making suggestions \cite{ford1999irfordecision}. To accomplish this goal,  a retrieval system should be aware of the user context in which the information need is to be placed \cite{xu2023foreword}. 
For complex decision-making tasks, the system is expected to scaffold the task step-by-step \cite{borlund2013interactive}. The system should also provide transparent retrieval process and explainable results to provide reasonable decision-making supports \cite{yin2023trustworthy}. Besides, understanding how people make decisions is also important. Collaboration with other communities will likely be necessary to make progress in decision understanding.  These communities may include human-computer interaction, behavioural economics, psychology, cognitive science, as well as specific domain communities like the clinical communities \cite{tsvetkov2015cognitive, ingwersen1984psychological}. 
Finally, placing IR into LLMs to provide AI-Generated Actions (AIGA) is also an emerging topic for the extension of IR \cite{janner2023deepgenerative}. 

\paragraph{Going beyond Cyberspace?}
IR systems have conventionally emphasized digital content within the realm of cyberspace.
However, advancements in human-computer interaction~\cite{esposito2021biosignal}, unmanned vehicles~\cite{mohsan2022towards}, and robotics~\cite{zhu2022soft} enable us to envision the retrieval of contents in the physical world.
For instance, the extended IR system may explore information in the physical world such as the wind speed around and the source of noise. Further extensions may incorporate the search and delivery of physical items. These extensions build on the aforementioned ranking-generation hybrids, human-machine hybrids, and finding-decision hybrids. Moreover, new concepts around interaction interfaces, system architecture, and feedback mechanisms will arise. 
Towards these ends, we may open up more cross-domain research directions and entangle IR with various emerging techniques in other fields such as Metaverse~\cite{mystakidis2022metaverse}
and Embodied Artificial Intelligence~\cite{shenavarmasouleh2022embodied}. 
As a result, there will be many research opportunities regarding the development, operation, maintenance, and regulation of such IR systems, increasing the value of the IR industry remarkably.


\section{Large Language Models for IR}
\label{Sec:big_model_IR}
IR systems have long been divided into five major components: user modeling, indexing, matching/ranking, evaluation, and user interaction.
By pre-training on large-scale corpora and fine-tuning to follow human instructions, LLMs have demonstrated superior capability in language understanding, generation, interaction, and knowledge reasoning~\cite{Ouyang2022TrainingLM}.
Therefore, it is reasonable to anticipate that LLMs will considerably augment the major IR components from various perspectives. In this section, we discuss the possibilities of LLMs for each major IR component.  

\subsection{User Modeling}
\label{sec:usermodel}

User modeling aims to accurately represent users and their information needs by understanding their characteristics, preferences, and behaviors~\cite{kis2001,Interact1994,websearch2002,attentive2019}. 
In view of the acknowledged capabilities of LLMs, we enumerate several potential directions that LLMs can enhance user modeling.

\begin{itemize}
\item \textbf{Language Understanding}. Most basically, LLMs can enhance the understanding of user queries, enabling more accurate analysis and leading to more relevant search results.

\item \textbf{Behavior Understanding}. LLMs allow the analyses of user behaviors in the semantic level of item content, offering a comprehensive understanding of preferences and behaviors. By analyzing data like click-stream, search log, and interaction history, models can identify patterns and relationships, resulting in more accurate user models.

\item \textbf{Personalization}. LLMs can build comprehensive user models, incorporating various characteristics and preferences. By analyzing social media activity, online behavior, and integrating with IoT devices, models can consider the factors like physical environment and emotional states, leading to more personalized recommendations.

\item \textbf{Conversational Interfaces}. LLMs faciliate more natural and seamless interactions, generating context-aware responses. Conversational interfaces can incorporate sentiment analysis and emotional response generation, thereby enhancing personalized and engaging user experiences.

\item \textbf{Hybrid Modeling}. Combining LLMs with rule-based or collaborative filtering models can yield stronger hybrid models. Such hybrid modeling unifies the strengths of different approaches, improving the personalization and accuracy of search and recommendation~\cite{tallrec,chat-rec}.
\end{itemize}

The advancement of LLMs is expected to significantly enhance the user modeling in IR systems. Nonetheless, certain challenges, such as data privacy, biases in data and models, and the need for extensive training data, need to be tackled. Addressing these challenges is imperative to fully exploit the potential benefits of LLMs for user modeling in IR, while mitigating any potential negative effects.

\subsection{Indexing}
\label{sec:indexing}

The emergence of LLMs provides the basis for generative retrieval, a new retrieval paradigm where the indexing component is dramatically changed. Recently, dense retrieval has been extensively studied and the approaches based on the standard MIPS index and nearest neighbor search are common \cite{karpukhin2020dense}. Given the success of Transformers as a good associative memory store or search index, Tay et al. \cite{tay2022transformer} propose a novel architecture called differentiable search index (DSI) where the index is stored in the model parameters. DSI uses LLMs to directly learn the mapping of queries to relevant document IDs. It enables the model's internal memory to act as an index, thus greatly simplifying the entire retrieval process. 
Inspired by the work of DSI, we identify some directions that LLMs will change the technology of indexing:

\begin{itemize}
    \item \textbf{From Static to Dynamic.} 
     Indexing systems based on LLMs need be dynamic, as all corpus information is encoded within the LLM parameters. Incremental index updates are a specific instance of model updates, as noted in the study by \cite{sun2020test}.
    
    \item \textbf{From Keyword-based to Semantics-oriented.} 
    Indexing systems based on LLMs should be semantic. 
    Thanks to the powerful contextual modeling capabilities of LLMs, indexing systems based on LLMs are able to find the documents that are related to the query in a more nuanced way.
    
    \item \textbf{From Uni-modal to Multi-modal.} 
    Indexing systems utilizing LLMs have the potential to be multi-modal. The development of multi-modal LLMs will facilitate the indexing systems capable of indexing various modalities of data in a unified manner, including but not limited to texts, images, and videos.
\end{itemize}

\subsection{Matching/Ranking}
\label{sec:matching}

LLMs have demonstrated remarkable capability to understand and rank complex content, including both single-modal and multi-modal data. In this part, we focus on two topics: (1) If generative models can already provide exact answers, is ranking still necessary? (2) What are the future research directions in the ranking, and what problems remain to be solved?

While generative LLMs can provide coherent answers to user queries, users still need to know which document, image, or webpage is most relevant to their query, so as to verify the answer generated by the LLM. The ranking results from a retrieval system can improve the interpretability of LLM, and provide the trustworthy information to support the answer generation. The retrieval system typically acts as a plugin to LLMs \cite{Feng2023KnowledgeRV,qin2023webcpm}, providing knowledge acquisition abilities.

In future research, LLMs offer many interesting directions to explore in ranking. Under the generative paradigm, we should prioritize ranking and improve the ranking of results to ensure a good user experience and high satisfaction. This is also a more essential goal of ranking in IR. When ranking search results, we should focus on the integrity of the returned results. Instead of simply returning a ranking list based on relevance, the model should return the information that is more integrated and relevant to users' real needs.

In terms of ranking evaluation, existing methods are designed for the form of returning a list of documents, which is no longer suitable under the generative paradigm. There are promising directions for future research in the field of ranking. An example is to leverage the LLMs as the human simulator to measure the user satisfaction and experience of a ranking method \cite{Sun2022MetaphoricalUS}.

\subsection{Evaluation}
\label{evaluation}

For validate the effectiveness of LLM-enhanced IR approaches, it is essential to develop proper metrics and datasets. 
Conventional IR metrics, such as Precision and Recall, Mean Reciprocal Rank (MRR, \cite{craswell2009mean}), Mean Average Precision (MAP, \cite{zhu2004recall}), Normalized discounted cumulative gain (nDCG, \cite{jarvelin2002ndcg,sun2023chatgpt}), still play a critical role in IR model evaluation.  However, with the wide application of LLMs in IR, tailored evaluation strategies become essential to justify the effectiveness of LLMs. 
To this end, some characteristics needed to be emphasized including: 
\begin{itemize}
\item \textbf{Robustness}.  Many models are sensitive to distributional differences between training and test data \cite{mitra2017neural,zhan2022evaluating}. It is eager to see the generalization of LLM-enhanced IR models in out-of-distribution (OOD) scenarios. 
\item \textbf{Interpretability}. Neural IR models rely on dense document representations, and suffer from poor interpretability, which is different from previous sparse IR models (e.g.,  BM25) with explicit term matching \cite{llordes2023explain}.
\item \textbf{Efficiency}. LLMs may require additional training or fine-tuning to adapt to IR tasks, and storage is also a severe bottleneck in the training scenarios \cite{santhanam2022moving,he2022rethinking}.
\item \textbf{Reliability}. LLMs are capable of directly retrieving knowledge from the model itself and generating results for users. However, they are vulnerable to adversarial inputs and some small character changes would negatively affect its reliability \cite{shen2023chatgpt}. 
\end{itemize}

\subsection{Interaction}
\label{sec:interaction}

The interaction between users and traditional search engines typically involves three steps \cite{baeza1999modern}. First, a user submits a query generally expressed as keywords. Then, the search engine processes the query, retrieves its index and presents the results in the form of a list. Finally, the user browses the results and clicks on the most relevant links to the query. 

With the development of conversational technology, conversational search has become a new retrieval paradigm \cite{ radlinski2017theoretical, ren2021conversations}. The retrieval process has become an iterative form of multiple rounds of dialogue. A user initiates the conversation by asking a question in natural language. The search engine processes the user's input, retrieves its index and generates a response. Then, the user may provide additional information or ask follow-up questions, and the search engine continues to process the user's input and generate responses until the user is satisfied. 

LLMs \cite{liu2023summary} have the potential to overturn the interaction between users and IR systems. For example, search engines powered by LLMs can serve as an AI assistant, such as Windows Copilot, Bing Chat. It can generate responses based on the conversation's context and the user's needs in real-time, without relying on pre-programmed responses. 
As long as the user inputs a question, it can help retrieve information, search knowledge, use Apps, and call plugins in the simplest way \cite{schick2023toolformer}. 
The IR system powered by LLMs possesses the capability to comprehend and address complex queries, thereby enhancing the interaction process in terms of intuitiveness, personalization, efficiency, effectiveness, responsiveness, and friendliness.

\section{IR for Large Language Models}
\label{Sec:IR_for_big_model}

LLMs exhibit certain inherent limitations that are commonly encountered in generative language models. Firstly, they may occasionally generate erroneous or nonsensical responses, a phenomenon referred to as "hallucinations"\footnote{Introducing ChatGPT, \url{https://openai.com/blog/chatgpt}}. Secondly, they are trained on fixed corpora, which restricts their ability to answer questions that require newly emerged knowledge or information after the training date~\cite{komeili2021internetaugmented,lazaridou2022internetaugmented}; the stored knowledge within the model can be inevitably incomplete, outdated, or even incorrect \cite{he2022rethinking}. Thirdly, they can hardly respond to ``private'' questions that require access to confidential data sources, which are not available during the training of LLMs.



Utilizing the retrieval capabilities of IR models presents a viable approach for addressing the above limitations of LLMs. By incorporating retrieval, we can leverage relevant knowledge from external knowledge bases  during the generation process, thereby reducing the occurrence of hallucinations. Retrieval models possess strong ability to access up-to-date information, enabling LLMs to respond with fresh and relevant information. Moreover, the combination of retrieval and LLMs is particularly valuable in data-sensitive scenarios, where internal data cannot be utilized for language model training, factual accuracy is of utmost importance, or the retrieval pool may vary over time~\cite{lee2022factuality, petroni2019language}.

In this section, we discuss three directions that augment LLMs with retrieval in different phrases: pre-training, fine-tuning, and leveraging black-box LLMs such as ChatGPT that only offer APIs.  

\subsection{Pre-training LLMs with Retrieval}


There is limited work on pre-training LLMs that incorporate a retrieval module. 
One notable example is Atlas~\cite{izacard2022atlas}, which pre-trains an encoder-decoder T5 with an extra retrieval module and demonstrates its ability on knowledge-intensive tasks with very few training samples.
REALM enhances encoder-only language model training by incorporating a neural knowledge retriever that extracts information from a text-based knowledge database \cite{guu2020realm}.
RETRO is a retrieval-augmented decoder-only language model, where a chunked cross-attention module is employed to aggregate retrieved text from a retrieval pool containing trillions of tokens \cite{borgeaud2022improving}.

In a recent study, based on the auto-regressive language model RETRO, Wang et al. \cite{wang2023shall} studies the question whether it is useful to pre-train LLMs with retrieval. 
As a result, the pre-trained retrieval-augmented language model RETRO outperforms the vanilla GPT on text generation tasks with higher factual accuracy, and on knowledge-intensive tasks. In addition, in open-domain question answering tasks, RETRO largely outperforms GPT which just incorporates retrieval at the fine-tuning stage.
Note that there is a trade-off between the computation overhead from using retrieval and the performance. Wang et al. \cite{wang2023shall} suggest a flexible implementation that specifies the number of tokens to generate with the current retrieval result before the next retrieval.

In addition to incorporating a retrieval module into the pre-training of LLMs, it is also valuable to investigate how retrieval functionalities can enhance the pre-training process itself. 
Generally, the retrieved evidence serves as the \emph{privileged information} within the input context. Such privileged information is often costly or impractical to obtain during the inference stage, though it is available during the training phase \cite{xu2020privileged}. By having both the retrieved evidence and the processed output answer (or subsequent tokens) accessible during training, it is possible to pre-train (or simply fine-tune) LLMs to improve their performance on knowledge-intensive tasks \cite{nakano2021webgpt,schick2023toolformer}.

\subsection{Fine-Tuning LLMs with Retrieval Adapters}\label{subsec:finetune}
Pre-training or fine-tuning the entire LLM with retrieval could enhance its retrieval capabilities. However, the considerable cost makes it prohibitive to use. Considering that LLMs serve as a foundation for downstream tasks, updating all parameters should be approached cautiously and infrequently. 
Therefore, an alternative is to fine-tune LLMs with search adapters, which is more efficient and cost-effective. Adapters are plug-and-play modules for LLM with only a small number of parameters. 
They enable LLMs to acquire task-specific capabilities without affecting the original parameters, while still achieving comparable or superior performance compared to updating the entire models.

There have been several studies on LLM adapters, which can be roughly classified into two categories: token-based methods and layer-based methods~\cite{he2022towards,hu2023llm}.
The former seeks to insert task-related anchor tokens into the input sequence for fine-tuning~\cite{li-liang-2021-prefix,liu-etal-2022-p,liu2021gpt}.
For example, Prompt tuning prepends several additional task-specific tunable tokens into the input sequence~\cite{lester-etal-2021-power}.
The latter seeks to insert additional layers into the models and fine-tune these layers only~\cite{houlsby2019parameter,pfeiffer-etal-2021-adapterfusion,zhang2023adaptive}.
For example, Hu et al. \cite{hu2021lora} propose LoRA which adds a trainable low-rank dense layer before the self-attention in transformers.
LLMs with adapters have been demonstrated effective on some downstream tasks such as machine reading comprehension.

Nevertheless, the following research questions remain to be explored to empower LLMs with retrieval capabilities through adapters.
\begin{itemize}
\item What adapters are best suited for retrieval? What neural network modules are best suited for retrieval adapters?
\item How to fine-tune the retrieval adapters? 
Which fine-tuning techniques can promote the retrieval capability in highest degree without hurting the inherent capabilities of LLM?
\item Why do retrieval adapters work? How do we know the capability boundary of retrieval adapters, and what can we do to avoid potential failures?
\end{itemize}

Fine-tuning LLMs with retrieval adapters can provide benefits in various aspects.
On one hand, it offers new opportunities for researchers with limited computation resources.
Progress in scientific research cannot be achieved without the participation of researchers around the world.
On the other hand, the lower cost of search adapters allows for wider applications, especially when data privacy is a top priority.
Small-size institutions can have LLMs equipped with their own retrieval systems without exposing the raw data.

\subsection{Augmenting Black-box LLMs with Retrieval}
In many scenarios, LLMs can only be accessed through remote APIs, and the possibility of fine-tuning these models is restricted. The most prevalent approach to leveraging LLMs is treating them as black-box systems and using customized prompts that integrate evidence from external data sources. This method enables users to obtain desired outputs from the LLMs by providing tailored prompts that incorporate relevant external evidence.

There are some preliminary studies on this interesting topic. shuster et al. \cite{shuster2022language} proposed a modular system SeeKeR which searches for and chooses knowledge with internet search as a module during language generation. Similarly, Komeili et al. \cite{komeili2021internetaugmented} and Lazaridou et al. \cite{lazaridou2022internetaugmented} also introduced approaches which condition on the results from internet search engines (such as Google Search) to generate a response. 
He et al. \cite{he2022rethinking} proposed a post-processing approach named ``rethinking with retrieval (RR)'' to solve the same problem. RR uses the chain-of-thought (COT) prompting to generate multiple reasoning paths, then retrieves external evidences based on the steps in these paths. Experiments on three complex reasoning tasks and different datasets demonstrate that RR outperforms all baselines without additional pre-training or fine-tuning. Ram et al. \cite{ram2023incontext} introduced an in-context Retrieval-Augmented Language Modeling (RALM) method that simply uses off-the-shelf general purpose retrievers to retrieve documents and preprends retrieved results to the input of language models. Peng et al. \cite{peng2023check} proposed a system named LLM-Augmenter, which augments a fixed black-box LLM with a set of plug-and-play modules. Given a query, LLM-Augmenter first retrieves evidences from external data sources, then generates a prompt that contains the retrieved evidences for ChatGPT. Their experiments on the information seeking dialog and open-domain wiki question answering tasks show this method could significantly reduce the hallucination problem of LLMs.

Recently, OpenAI officially released the Retrieval Plugin\footnote{ChatGPT Retrieval Plugin, \url{https://github.com/openai/chatgpt-retrieval-plugin}}, which provides semantic search and retrieval functionality of personal or organizational documents. Relevant documents snippets can be retrieved from external sources, such as personal emails or internal organizational files, with this plugin. 

There are several key components in the pipeline of augmenting LLMs with retrieval, which are worth to study in the future. 
\begin{itemize}
    \item \textbf{Design of Retriever}. Intuitively, a better retriever could return results with higher quality, and thereby helps to generate better response. Should we use a dense retriever or a sparse retriever in generating grounding documents? Which kind of information to index, documents or passages?
    \item \textbf{Context Modeling}. How to generate an explicit keyword query or representation vector that can describe the current information need and retrieve useful results for language generation? ChatGPT-like LLMs are optimized for dialogue rather than search, hence how to derive search intent from the multi-turn chat history remains a challenge.
    \item \textbf{Selection of Grounding Documents}. In addition to pure relevance, other factors of diversity, informativeness, and freshness should also be considered when selecting a small set of documents.
    \item \textbf{Prompting Mechanism}. It remains a challenge to generate good prompts that can help improve generation quality with the given grounding documents. 
\end{itemize}


\section{LLMs + IR: New Paradigm and Framework}
\label{Sec:big_model+IR}

Traditional IR models are designed to meet human information needs through the interactions between users and systems. 
IR models do not preserve knowledge, instead, they search for external knowledge or information to meet the information needs efficiently. 
With the emergence of LLMs, we propose a new technical paradigm of IR: as shown in Figure~\ref{fig:1}, the key elements are LLMs, IR models and humans.
\begin{itemize}
    \item LLMs provide valuable (internal) knowledge and information to meet human information needs, and their reasoning capacity makes it easier to provide high-quality responses.
    \item IR models that search for external knowledge are indispensable, which provide up-to-date and relevant information for both LLMs and humans. 
    While LLMs may suffice in directly generating answers to meet human information needs for simple questions, the significance of IR models becomes more pronounced when dealing with complex and difficult problems.
    \item Humans raise information needs in the paradigm and are the tutor for both LLMs and IR models. They endow the system with human values and behavioral characteristics, making it serve users better.
\end{itemize}
The synergistic relationship among them not only facilitates mutual enhancement but also enables the fulfillment of new tasks as an organic whole.

\subsection{Importance of the Three Modules}
In accordance with the new IR technical paradigm, an important question arises: Are all three elements essential for this new paradigm? In the following subsection, we will provide our insights and responses.

\subsubsection{Paradigm without LLMs}
Without LLMs, the paradigm degrades to traditional IR, accompanied by inherent challenges:
\begin{itemize}
    \item \textbf{Dependency on the Internet}. As IR models do not retain knowledge or information themselves, they rely on the Internet to acquire external knowledge, potentially limiting their applicability in certain scenarios.
    \item \textbf{Lacking reasoning ability.} Existing IR models mainly provide collected knowledge/information to fulfill human information needs, lacking the ability to assist users in comprehending the information. Better reasoning ability will deliver more user-friendly and valuable results for humans.
\end{itemize}

\subsubsection{Paradigm without IR}
In the era of LLMs, the presence of an IR system is critical as it addresses the limitations of generative language models. Without an IR system, LLMs encounter the following challenges:
\begin{itemize}
    \item \textbf{Lacking factual consistency}.
One inherent limitation of generative LLMs is the lack of factual consistency, resulting from their training data and the potential generation of false information. In contrast, the IR system frees from this issue by storing and offering factual information through keyword matching, thereby complementing the shortcomings of LLMs.

    \item \textbf{Lacking effective integration of new and existing knowledge.}
Large Language models have the capacity to learn knowledge from large-scale data, such as commonsense information. However, relying solely on LLMs may neglect the crucial mechanism of effectively and efficiently integrating new and existing knowledge. In the human brain, memory involves a complex network of interconnected brain regions, including the hippocampus, prefrontal cortex, and other cortical areas. Retrieval processes are closely linked to memory consolidation and the reactivation of neural networks associated with previously encoded information. When we retrieve information from memory, this activity can strengthen the neural connections related to that information, making it more accessible for future retrieval and improving overall memory performance.
    
\end{itemize}

\subsubsection{Paradigm without Human}
The absence of human input and feedback hinders both LLMs and IR models from delivering personalized information services. The associated challenges are:
\begin{itemize}
    \item \textbf{Identical information services.}
Large-scale dialogue systems use a generation-based approach to provide information to users. The generated content is not personalized, which can cause the issues that the content does not match the user's real intention.
    \item \textbf{Uncontrollable value of society.}
Without human feedback, LLMs are unable to account for user personality and values, resulting in information services that do not adequately align with societal values.
\end{itemize}

\subsection{Summary}

The rise of dialogue-based LLMs, exemplified by platforms like ChatGPT from OpenAI and New Bing from Microsoft, has sparked a trend that could potentially replace traditional IR systems. However, as previously discussed, in the era of LLMs, human information needs and feedback, along with the continued relevance of traditional IR models, remain vital components in the development of trustworthy information service systems.

\section{Challenges and Future}
\label{Sec:challenges_discussions}
While LLMs for IR hold promise, they also present numerous challenges and unanswered questions. In the final section of this article, we discuss some selected issues to outline future directions.

\begin{itemize}
    \item {\bf High Computational Costs}. The primary challenge in using LLMs is their high computational cost. This poses a significant barrier for small and medium-sized research laboratories and companies, hindering their integration of LLMs into daily workflows and products.  Even large companies with ample computational resources face cost pressures when deploying LLMs for online search, recommendation, and advertisement services due to the immense volume of user requests. Common solutions include compressing LLMs, reducing their size from hundreds of billions to tens billions or even smaller, especially before online deployment. Additionally, efforts to develop more efficient and cost-effective hardware for training and inference are underway to address the cost challenge.
    
    \item {\bf General-purpose v.s. Domain-specific}.   
    LLMs have demonstrated impressive capabilities in general-purpose tasks like text generation and chatting, owing to their pre-training and fine-tuning on large-scale Internet corpora. However, it is widely recognized that LLMs face limitations when it comes to adapting to domain-specific tasks. One one hand, high-quality professional domain knowledge, which is often not abundantly available on the Internet, makes it prohibitive to pre-train and fine-tune LLMs. On the other hand, domain-specific knowledge is not always expressed in natural language; it may be represented as semi-structured or structured tables, heuristic rules, equations, and more. Enabling LLMs to effectively handle domain-specific tasks is crucial not only for the specific domains themselves but also for enhancing the overall capabilities and applications of LLMs.
\item {\bf Trustworthiness}. There is a widely acknowledged concern that LLMs currently lack the ability to provide reliable and trustworthy answers to user queries. While LLMs can generate explanations and cite sources, it has been observed that a significant portion of these explanations and citations are illogical, inappropriate, or even fake. This poses a substantial risk in real-world search and recommendation scenarios, as generating misleading explanations, answers, and information sources can have detrimental effects on the community at large. To address this issue and enhance the trustworthiness of LLMs, it is crucial to enable LLMs to have a clear understanding of their knowledge and limitations. One potential solution is allowing LLMs to decline providing an answer when uncertain.

\item {\bf Controllable Generation}.Considering the public nature of search engines and recommendation systems, it is important to address regulatory and ethical considerations such as fairness, impartiality, and human values when presenting content to users. While LLMs demonstrate proficiency in generating text, they often lack a deep understanding of the meaning behind the generated words. Ensuring that the generated content meets the necessary regulatory and ethical requirements remains a significant challenge, and has no effective solutions for now.

\item {\bf High-quality Data:} High-quality data plays a vital role in the development and improvement of LLMs. The success of LLMs heavily relies on the continuous provision of human-labeled data. It is crucial that the labeled data not only meets a certain quantity threshold but also maintains high quality. Obtaining high-quality data in real-world application scenarios involves multiple steps such as data cleaning, data labeling, and data quality evaluation. Professional data annotation providers play a crucial role in supporting these processes. Additionally, it is essential to develop advanced, professional, and sustainable data annotation approaches to meet the growing demand for high-quality data in LLM applications.

\item {\bf Long-context Dependency}. Existing LLMs have limited capacity to handle long contexts, while IR tasks rely on long-term context to effectively capture and understand user intent. It is crucial to enable LLM-enhanced IR systems to model users' long-term intent that spans a large range of period.

\item {\bf Serving Time Requirements}. The latency in serving LLM results significantly lags behind the time requirements of information retrieval (IR) systems. This presents efficiency challenges when integrating LLMs into IR, thereby impacting the online user experience.

\item \textbf{Presentation Format}. Traditional IR systems present ranked lists of content, while LLMs excel at generating new information. How to design a new presentation format that effectively fulfills user needs in LLM-enhanced IR remains an open question.

\item \textbf{Integrating Structural Information}. LLMs primarily rely on textual sequence information, whereas IR systems require the integration of structural information such as user-item interactions and web linkage data. Effectively leveraging this structural information in LLM-based IR systems is an unresolved issue.

\item \textbf{Balance between Generative and Retrieved Data}. LLMs leverage deep learning and reinforcement learning to generate content at scale, but their generated content may have limitations in terms of freshness and credibility. In contrast, retrieval can provide the content from the Web, ensuring the latest information. Balancing these two types of data in real-life applications is a significant challenge for improving overall performance. One approach is to refine user needs and classify them into different groups, allowing for appropriate data generation methods or balancing ratios. In addition, retrieval can be used to provide additional information to enhance content generation, or assist in screening and filtering the generated content for information grounding.

\item \textbf{Content Quality and Credibility}. While LLMs are effective in generating content, they can also produce low-quality or even misinformation-filled content. The proliferation of such content on the Internet can disrupt the existing data ecology and impact applications like search engine and recommendation system. Traditional quality assessment techniques like PageRank may not be effective in this context.
It is difficult for existing techniques to identify low-quality or misleading content. In the era of AI-generated content, new mechanisms are needed to assess data quality and distinguish between generated and reliable content. One approach is manual or community review to ensure content accuracy, but it is time-consuming and does not scale well. Another approach is leveraging LLM-powered machine learning techniques to train models that can recognize AI-generated content and evaluate its quality. These models can then be applied to select, label, or filter content for different applications.

\item \textbf{Content Creation Environment}. The proliferation of generated content introduces challenges for content creators, and may even reshape the content ecosystem. The presence of generated content intensifies competition within the content market, pushing content creators to continuously enhance their writing quality, foster innovation, and proactively adapt to industry changes. Moreover, it may influence users' perception of value and demand for content, prompting content creators to constantly adjust their writing approach and strategies. Despite the challenges, LLMs also provide new opportunities for content creators to collaborate and develop effective platforms that facilitate more productive and higher-quality content generation.

\end{itemize}

\bibliographystyle{elsarticle-num}
\bibliography{sigirforum}

\end{document}